\documentclass[aps,preprint,nofootinbib,showpacs]{revtex4}
\usepackage{amsfonts}
\usepackage{amsmath}
\usepackage{amssymb}
\usepackage{graphicx}%

\newcommand{\beq}{\begin{equation}}
\newcommand{\eeq}{\end{equation}}
\newcommand{\beqn}{\begin{eqnarray}}
\newcommand{\eeqn}{\end{eqnarray}}
\newcommand{\bearr}{\begin{array}}
\newcommand{\enarr}{\end{array}}

\newcommand{\toref}[1]{\mbox{(\ref{#1})}}

\newcommand{\eps}{\varepsilon}

\begin{document}


\title{From synchronization to Lyapunov exponents and back}

\author{Antonio Politi$^1$, Francesco Ginelli$^2$, Serhiy Yanchuk$^{3,4,5}$,
Yuri Maistrenko$^{3,6}$} 

\affiliation{$^6$ CNR - Istituto dei Sistemi Complessi, Via Madonna del Piano,
10, I-50019 Sesto Fiorentino, Italy\\
$^2$ CEA -- Service de Physique de l'Etat Condens\'e, Centre 
d'Etudes de Saclay, 91191 Gif-sur-Yvette, France\\
$^3$ Institute of Mathematics, National Academy of Sciences of Ukraine,
01601 Kyiv, Ukraine\\
$^4$ Weierstrass Institute for Applied Analysis and Stochastics,
Mohrenstrasse 39, 10117  Berlin, Germany\\
$^5$ Institute of Mathematics, Humboldt University of Berlin, Unter den
Linden 6, 10099 Berlin, Germany\\
$^6$ Institute of Medicine and Virtual Institute of Neuromodulation,
Research Centre J\"ulich, 52425 J\"ulich, Germany}

\date{\today}
\vskip 2.cm
\begin{abstract}
The goal of this paper is twofold. In the first part we discuss a general
approach to determine Lyapunov exponents from ensemble- rather than
time-averages. The approach passes through the identification of locally stable and
unstable manifolds (the Lyapunov vectors), thereby revealing an analogy with
generalized synchronization. The method is then applied to a periodically
forced chaotic oscillator to show that the modulus of the Lyapunov exponent 
associated to the phase dynamics increases quadratically with the coupling 
strength and it is
therefore different from zero already below the onset of phase-synchronization. 
The analytical calculations are carried out for a model, the generalized special
flow, that we construct as a simplified version of the periodically forced
R\"ossler oscillator.

\end{abstract}

\maketitle

\section{Introduction}

As soon as synchronization phenomena in chaotic systems have been 
discovered \cite{FY83, Pikov84}, the standard tools of nonlinear dynamics have
been implemented in order to clarify this phenomenon. This is particularly true
for the Lyapunov exponents (LEs), \cite{Oseledec} because they measure the
degree of stability along different directions and are thus the natural
candidates to quantify the degree of synchronization of different regimes.
However, several subtleties have been immediately discovered. For instance, the
negativity of the ``transversal'' LE is only a necessary condition for the
stability of complete synchronization: (i) in low-dimensional systems,
fluctuations of the finite-time LEs may render the synchronized regime unstable
even when the ``average" exponent is negative\cite{R1}; (ii) in high
dimensional systems, it has been ascertained that the propagation of
finite-amplitude perturbations can sustain an unsynchronized
regime, in spite of its linear stability\cite{BLT01, Pikov02}. A still open
problem concerns the behaviour of the LEs in the context of
phase synchronization \cite{Pikov96a, Parlitz96} and more precisely of the
exponent quantifying the stability of the phase dynamics. In fact, it is often
claimed that this LE is the right order-parameter to characterize the
onset of phase-synchronization: below the transition it is conjectured to be
zero, while it is strictly negative above the transition \cite{Pikov96a, Hu03}.
However, the situation is certainly less simple than initially believed because
a negative exponent has been found also in correspondence of locking
phenomena occurring below the onset of phase-synchronization\cite{Kye03}.
It is therefore important to clarify analytically the stability of the dynamics
along the ``phase'' direction: in the absence of coupling, this is a marginally
stable direction and it is thus natural to expect some difficulties. Here, we
develop a method that allows concluding that the LE corresponding to the phase
dynamics is different from zero (and possibly positive) as soon as the coupling
is switched on and therefore even below the onset of phase-synchronization.

One of the main problems is the lack of analytical methods for determining
even perturbatively the LEs. Some ideas have been put forward for the maximum
exponent \cite{R2,R3}, because almost any initial condition eventually grows with
the maximum rate and no special care is required to tune the direction of the
perturbation. However, very little is known for the other exponents, starting
already from the second one. This is precisely what is needed to determine the
stability of phase dynamics in the simplest system exhibiting phase
synchronization, i.e. in a periodically forced chaotic attractor, where the
first LE accounts for the overall instability of the chaotic dynamics. Here, we
attack and solve the problem by developing a formalism to determine
LEs as ensemble- rather than time-averages. Similar ideas have been already
discussed by Ershov and Potapov \cite{ErshovPotapov}, although they have not
gone much beyond the level of formal statements. In fact, their method relies
on the determination of the growth rates of hypervolumes of increasing
dimension. While this idea proved very effective for the development of a
powerful algorithm to compute the LEs\cite{Lyapunov}, its ensemble-average
extension has some limitations due to the difficulty of disentangling the
various exponents. The advantage of our approach is that we are able to
associate each non-degenerate LE to a field of local directions, the
{\it Lyapunov vectors} (LVs). Roughly speaking, the $i$th LV is determined into
two steps: the first one consists in iterating forward in time a hypervolume of
dimension $i$ in tangent space to identify the local orientation of the most
expanding $i$ directions (this is also considered in \cite{ErshovPotapov}); the
second step consists in iterating backward a vector lying within such a
hypervolume. As a result, a coordinate-independent LV can be determined: the LE
is finally obtained by averaging the corresponding instantaneous
expansion/contraction rate over the entire phase-space, according to the
invariant measure.

An objective identification of LVs is particularly interesting in the study of
the hydrodynamic behaviour of extended systems. In the last years, mostly as a
consequence of the pioneering work of Posch and collaborators \cite{Posh1,Posh2},
it has been discovered that in models of fluids (more in general in Hamiltonian
systems) the directions corresponding to the smallest (in absolute value) LEs
almost coincide with long-wavelength Fourier modes. This observation has in turn
suggested that the Lyapunov analysis naturally leads to a hydrodynamic
description without the need of introducing a suitable coarse graining. However,
progress has been hindered by the lack of an absolute definition of the
Lyapunov ``modes", that have been mostly identified with the vectors arising
from the implementation of the Gram-Schmidt orthogonalization procedure
during a standard computation of the LEs.
The only examples of a philososphy similar to that one outlined in the
present paper concern the chronotopic Lyapunov approach \cite{PLT} and the
characterization of space-time chaos \cite{Chate}. 

It is also interesting to notice that the problem of identifying the LVs is
itself equivalent to a problem of (generalized) synchronization. In fact, the
Lyapunov vectors are determined by integrating a skew-product system composed of
the original nonlinear dynamics plus the ``forced" evolution in tangent space.
As a result, the direction of the LVs varies in a possibly singular way with
the position in real space. However, this difficulty does not hinder the 
LE determination, which results from an average that is substantially
insensitive to the presence of local singularities.

An analytic investigation of the stability of phase dynamics in a generic setup
is an extremely difficult task because of the lack of structural stability
of low-dimensional chaos. For this reason, it is convenient to consider suitable
simplified models. The simplest system where phase-synchronization has been
investigated is the so-called special flow\cite{Pikov97b}. This is basically a
skew-product system, where the phase dynamics is forced by the chaotic amplitude
dynamics. In this system, it is possible to estabilish analytically a certain
number of results, because the phase evolution is basically unidimensional and
there is no need to deal with the problem of identifying the direction of
perturbations. In order to perform a more realistic analysis of phase
synchronization, a suitable coupling between phase and amplitude dynamics has
been added to the special flow \cite{Pikov97}. Here, setting up a perturbative
approach for the weakly forced R\"ossler oscillator, we show that the structure
of the model proposed in Ref.~\cite{Pikov97} is quite similar to that one
expected in generic chaotic systems, whenever the presence of strong
dissipations allows eliminating the stable directions. Furthermore, in order to
simplify the analytic treatment of the LEs, we focus our attention on a
model that we call the generalized special flow (GSF), very similar to that
one analyzed in Ref.~\cite{Pikov97} but characterized by a finite Markov
partition. As a result, we find that the modulus of the second LE exponent 
increases
quadratically with the coupling strength and its corresponding smallness
justifies the claims often found in the literature that the second LE is equal
to zero below the onset of phase-synchronization. In other words, we conclude
that the LE is not the right order parameter to describe this transition.

More precisely, this paper is organized as follows. In the next section we 
introduce a general approach for the determination of Lyapunov exponents
through an average over the invariant measure. In section \ref{sec3} we present
our case-study model, the GSF, deriving it as a discrete--time approximation of
a periodically forced R\"ossler system. In sections \ref{LL} and
\ref{measure_lyap} we illustrate the perturbation expansion for the second LE,
the corresponding LV and the invariant measure. Finally some numerical results
are presented in section \ref{conc}, along with conclusions.

\section{A general approach for the determination of Lyapunov vectors and
Lyapunov exponents}
\label{sec2}
In this section we discuss a method to determine Lyapunov exponents
from suitable ensemble averages. It is easy to write down a formal meaningful
definition, but the problem lies in translating it into a workable procedure.
With reference to an $N$-dimensional discrete--time system, described by the
mapping rule
\begin{equation}\label{eq:start1}
{\bf x}_{t+1} = {\bf f}_d({\bf x}_t) \quad \quad {\bf x} \in {\mathcal R}^N  ,
\end{equation}
one can express the $i$th LE (as usual, LE are supposed to be ordered from the
largest to the smallest one) as
\begin{equation}
\lambda^{(i)} = \frac{1}{2} \int d{\bf x} P({\bf x})
\ln \left[
\frac {|| \partial_x{\bf f}_d {\bf V}^{(i)}({\bf x})||^2}
{||{\bf V}^{(i)}({\bf x})||^2}
\right]
\label{genLyap1}
\end{equation}
where
$P({\bf x})$ is the corresponding invariant measure, $\partial_x{\bf f}_d$ 
is the Jacobian of the transformation, and the Lyapunov vector 
${\bf V}^{(i)}({\bf x})$ identifies the $i$th most expanding direction in
$\bf x$.

With reference to a continuous--time system, ruled by the ordinary differential
equation
\begin{equation}\label{eq:start2}
\dot {\bf x} = {\bf f}_c({\bf x}) \quad \quad {\bf x} \in {\mathcal R}^N  .
\end{equation}
the $i$th LE writes as
\begin{equation}
\lambda^{(i)} = \int d{\bf x} P({\bf x}) 
\frac{ \big[ \partial_x {\bf f}_c {\bf V}^{(i)} ({\bf x}) \big]
  \bullet {\bf V}^{(i)}({\bf x})}
{||{\bf V}^{(i)}({\bf x})||^2}
\label{genLyap2}
\end{equation}
where $\bullet$ denotes the scalar product.

Unless a clear procedure to determine the LV is given,
Eqs.~(\ref{genLyap1},\ref{genLyap2}) are nothing but formal statements. As
anticipated in the introduction, ${\bf V}^{(i)}({\bf x})$ can be obtained by
following a two-step procedure. We start with a generic 
set of $i$ linearly independent vectors lying in the tangent space and let them
evolve in time. This is the standard procedure to determine LEs, and it
is well known that the hypervolume ${\bf Y}^{(i)}$ identified by such vectors
contains for, large enough times, the $i$ most expanding directions.
Furthermore, with reference to the set of orthogonal cordinates obtained by
implementing the Gram-Schmidt procedure, the component $v_k$ of a generic vector
$\bf v$ evolves according to the following differential equation (for the sake
of simplicity, we refer to continuous--time systems),
\begin{equation}\label{eq:gsf}
\dot v_k = \sum_{j=k}^i \sigma_{k,j}({\bf x}) v_j \qquad 1 \le k \le i
\end{equation}
where, as shown in Ref.~\cite{ErshovPotapov}, $\sigma_{k,j}$ does not
explicitely depend on time, but only through the position $\bf x$ in the phase
space. As a result, the $i$th Lyapunov exponent can be formally expressed
as the ensemble average of the local expansion rate $\sigma_{i,i}$, i.e.,
\begin{equation}
\lambda^{(i)} = \int d{\bf x} P({\bf x}) \sigma_{i,i}({\bf x})
\label{genLyap3}
\end{equation}
By comparing with Eq.~\toref{genLyap2}, one finds the obvious equality
\begin{equation}
\sigma_{i,i} = \frac{ \big[ \partial_x {\bf f}_c {\bf V}^{(i)} ({\bf x}) \big]
  \bullet {\bf V}^{(i)}({\bf x})}
{||{\bf V}^{(i)}({\bf x})||^2}
\label{genLyap4}
\end{equation}
In Sec. \ref{LL}, where this formalism is applied to a phase-synchronization problem,
we find that the only workable way to obtain an analytic expression for
$\sigma_{i,i}$ passes through the determination of the direction of the
corresponding LV vector ${\bf V}^{(i)}({\bf x})$.

Let us now consider the backward evolution of a generic vector
${\bf V}^{(i)} \in {\bf Y}^{(i)}$. Its direction is identified by the
$(i-1)$-dimensional vector
\begin{equation}\label{eq:direc}
{\bf u} \equiv (u_1,u_2,\ldots,u_{i-1})
\end{equation}
where $u_k = v_k/v_i$. From the definition of $\bf u$ and from 
Eq.~(\ref{eq:gsf}), one easily finds that the backward evolution follows the
equation
\begin{equation}\label{eq:lyap3}
\dot u_k = (\sigma_{i,i}-\sigma_{k,k}) u_k 
   -\sum_{j=k+1}^{i-1}\sigma_{k,j}(t) u_j - \sigma_{k,i}  \qquad 1 \le k < i
\end{equation}
This is a cascade of skew-product linear stable equations (they are stable
because the Lyapunov exponents are organized in descending order). The overall
stability is basically determined by the smallest $(\sigma_{k,k}-\sigma_{i,i})$
that is obtained for $k=i-1$. It is, therefore, sufficient to turn our attention
to the last ($i-1$) component of the vector $\bf V$. Its equation has the
following structure
\beq
\dot{u}(t) = \gamma u + \sigma(t)
\label{back1}
\eeq
where $\gamma = \lambda_i - \lambda_{i-1} < 0$ and we have dropped the 
subscript $i$ for simplicity. The value of the direction $u$ is obtained by
integrating this equation. By neglecting the temporal fluctuations of
$\gamma$ (it is not difficult to include them, but this is not important for
our final goal), the formal solution of Eq.~\toref{back1} reads
\beq
u({\bf x}(t)) = \int^t_{-\infty} e^{\gamma(t-\tau)} 
     \sigma({\bf x}) \,d\tau  \quad .
\label{back2}
\eeq
This equation does not simply tell us the value of $u$ at time $t$, but the value
of $u$ when the trajectory sits in $\bf x(t)$. It is in fact important to
investigate the dependence of $u$ on $\bf x$. We proceed by determining the
deviation $\delta_j u$ induced by a perturbation $\delta x_j$ of ${\bf x}$ along
the $j$th direction, 
\beq
 \delta_j u = \int^t_{-\infty} e^{\gamma(t-\tau)} \delta_j \sigma(\tau) \,d\tau 
 \label{back3}
\eeq
where, assuming a smooth dependence of $\sigma$ on $\bf x$,
(see below for a further discussion of this point),
\beq\label{eq:line1}
\delta_j \sigma(\tau) \approx \sigma_x(\tau) \delta x_j(\tau) =  \sigma_x(\tau) \delta x_j(t)
e^{\lambda_j (t- \tau)} \quad .
\eeq
(notice that the dynamics is flowing backward). If the Lyapunov exponent
$\lambda_j$ is negative, $\delta_j \sigma(\tau)$ decreases for $\tau \to
-\infty$ and the integral over $\tau$ in Eq.~\toref{back3} converges. As a
result, $\delta_j u$ is proportional to $\delta x_j$, indicating that the
direction of the LV is smooth along the $j$th direction. If $\lambda_j$ is
positive, $\delta_j \sigma(\tau)$ diverges, and below time $t_0$ where
\beq\label{eq:t0}
\delta x_j(t) e^{\lambda_j (t-t_0)} = 1
\eeq
linearization breaks down. In this case, $\delta \sigma(\tau)$ for $\tau <t_0$
is basically uncorrelated with its ``initial value" $\delta_j \sigma (t)$ and
one can estimate $\delta_j u$, by limiting the integral to the range $[t_0,t]$
\beq\label{back4}
 \delta_j u(t) = \delta x_j(t) \int^t_{t_0} d \tau e^{(\lambda_j+\gamma)(t-\tau)}
  \sigma_x(\tau)
\eeq
where $t_0$ is given by Eq.~\toref{eq:t0}. By bounding $\sigma_x$ with
constant functions and thereby performing the integral in 
Eq.~\toref{back4}, we finally obtain
\beq
\delta_j u(t) \approx \delta x_j(t) + \delta x_j(t)^{-\gamma/\lambda_j}
\label{back6}
\eeq.
The scaling behaviour is finally obtained as the smalles number between
1 and $-\gamma/\lambda_j$. If we now introduce the exponent $\eta_j$ to 
identify the scaling behaviour of the deviation of the LV direction when the
point of reference is moved along the $j$th direction in phase space, the
results are summarized in the following way

\begin{equation}
\label{bareta} 
\eta_j = \left\{
\begin{array}{lr}
1 \;\;&\mbox{ for } \lambda_j \le -\gamma  \\
-\gamma/\lambda_j  &\;\;\mbox{ for }\lambda_j > -\gamma
\end{array}\right.
\end{equation}
The former case corresponds to a smooth behavior (the derivative is finite),
while the latter one reveals a singular behaviour that is the signature of a
generalized synchronization.

Although most of the assumptions made to derive the above equation are quite
plausible (even though not rigorously proved), there is one point that needs
to be more carefully checked: the smoothness of $\sigma ({\bf x})$. In the
absence of a more careful analysis of this point, we can only claim that the
above equation provides an upper bound to the true range of smoothness for the
LV direction.

\section{From the periodically forced R\"ossler system to the generalized special
flow}
\label{sec3}

The first model where phase synchronization has been explored is the forced
R\"{o}ssler oscillator \cite{Pikov96a}. In this section we derive a discrete-time
mapping describing a forced R\"{o}ssler system in the limit of weak coupling.
We obtain what we call the Generalized Special Flow (GSF), because it extends a
mapping previously introduced to characterize the onset of phase
synchronization\cite{Pikov97b}.

The starting set of ordinary differential equations is
\begin{eqnarray}
\dot{x} & = & -y-z+\varepsilon y\cos(\Omega t+\psi_{0})\nonumber \\
\dot{y} & = & x+a_0y-\varepsilon x\sin(\Omega t+\psi_{0})\label{rossler}\\
\dot{z} & = & a_1+z(x-a_2)\nonumber \end{eqnarray}
where $\psi_{0}$ fixes the phase of the forcing term at time 0. It is convenient
to introduce cylindrical coordinates, namely $\mathbf{u}=(\varphi,r,z)$,
($x=r\cos\phi$, $y=r\sin\phi$). For the future sake of clarity,
let us denote with $\mathbf{S_{c}}$ the 3-dimensional space parametrized
by such coordinates. The differential equation (\ref{rossler}) writes
as
\begin{equation}
\dot{\mathbf{u}}=\mathbf{F}(\mathbf{u})+\varepsilon\mathbf{G}(\mathbf{u},\Omega t+\psi_{0})
\label{eqtot}
\end{equation}
where
\begin{eqnarray}
\mathbf{F} & = & \left[1+\frac{z}{r}\sin\phi+\frac{a_0}{2}\sin2\phi\,,\, a_0r\sin^{2}\phi-z\cos\phi\,,
  \, a_1+z(r\cos\phi-a_2)\right]\\
\mathbf{G} & = & \left[-\sin^{2}\phi\cos(\Omega t+\psi_{0})-\cos^{2}\phi\sin(\Omega t+\psi_{0})
\,,\,\frac{r}{\sqrt{2}}\sin2\phi\cos(\Omega t+\psi_{0}+\pi/4)\,,\,0\right]\nonumber
\end{eqnarray}
Note that system (\ref{eqtot}) can be written in the equivalent autonomous form
\[
\dot{\mathbf{u}}=\mathbf{F}(\mathbf{u})+\varepsilon\mathbf{G}(\mathbf{u},\psi),\quad
\dot{\psi}=\Omega,
\]
where $\psi$ denotes the phase of the forcing term.

We pass to a discrete-time description, by monitoring the system each time the
phase $\phi$ is a multiple of $2\pi$. In the new framework, the relevant
variables are $r$, $z$, and $\psi$, all measured when the Poincar\'{e} section is
crossed. The task is to determine the transformation mapping the state
$(r,z,\psi)$ onto $(r',z',\psi')$.

In order to obtain the expression of the map, it is necessary to formally
integrate the equations of motion from one to the next section. This
can be done, by expanding around the unperturbed solution for $\varepsilon=0$
(which must nevertheless be obtained numerically). The task is
anyhow worth, because it allows determining the structure of the resulting
map, which turns out to be (see appendix \ref{app1})

\begin{eqnarray}
\label{3dFlow}
\psi' & = & \psi+\langle T^{(0)}\rangle\,\Omega+A_{1}+\varepsilon\left(B_{1}^{c}\cos\psi+B_{1}^{s}\sin\psi\right)\nonumber \\
r' & = & A_{2}+\varepsilon\left(B_{2}^{c}\cos\psi+B_{2}^{s}\sin\psi\right)\label{eq:map}\\
z' & = & A_{3}+\varepsilon\left(B_{3}^{c}\cos\psi+B_{3}^{s}\sin\psi\right)\nonumber
\end{eqnarray}
where $\langle T^{(0)}\rangle$ is the average period of the unperturbed 
R\"ossler oscillator and 
$A_m$'s and $B_m$'s are functions of $z$ and $r$. As it is shown in
appendix \ref{app1}, they can be numerically determined by integrating the
appropriate set of equations. Up to first order in $\varepsilon$, the structure
of the model is fairly general as it is obtained for a generic periodically
forced oscillator represented in cylindrical coordinates (as long the phase of
the attractor can be unambiguously identified).

For the usual parameter values, the R\"{o}ssler attractor is characterized by a
strong contraction along one direction \cite{YMM}. As a result, one can neglect
the $z$ dependence since this variable is basically a function of $r$, and thus
write
\begin{eqnarray}
\label{2dFlow}
\psi' & = & \psi+\langle T^{(0)}\rangle\,
\Omega+A_{1}(r)+\varepsilon\left(B_{1}^{c}(r)\cos\psi+
   B_{1}^{s}(r)\sin\psi\right)\nonumber\\
r' & = & A_{2}(r)+\varepsilon\left(B_{2}^{c}(r)\cos\psi+B_{2}^{s}(r)\sin\psi\right)
\end{eqnarray}
where all the functions can be obtained by integrating numerically
the equations of motion of the single R\"{o}ssler oscillator.\footnote{Strictly
speaking, $A$ and $B$ functions in (\ref{eq:map}) and
(\ref{2dFlow}) are different (see appendix \ref{app1}). We
use the same notations here to simplify the presentation.} 

To simplify further manipulations, we finally recast equation \toref{2dFlow}
in the form
\begin{eqnarray}
\label{2dFlow2}
\psi' & = & \psi+K+A_{1}(r)+\varepsilon g_{1}(r)\cos\left( \psi+\beta_{1}(r)\right)\nonumber\\
r' & = & A_{2}(r)+\varepsilon\ g_{2}(r)\cos\left(\psi+\beta_{2}(r)\right)
\end{eqnarray}
where
\begin{eqnarray}
\label{transf}
B_i^c(r) & = & g_i(r)\cos \beta_i(r) \nonumber\\
B_i^s(r) & = & - g_i(r)\sin \beta_i(r)
\end{eqnarray}
for $i=1,2$. The parameter 
$K= \langle T^{(0)}\rangle\,\Omega -2\pi$ represents the detuning between the
original R\"ossler-system average frequency and the forcing frequency
$\Omega$.

\begin{figure}
\includegraphics[clip,scale=0.5]{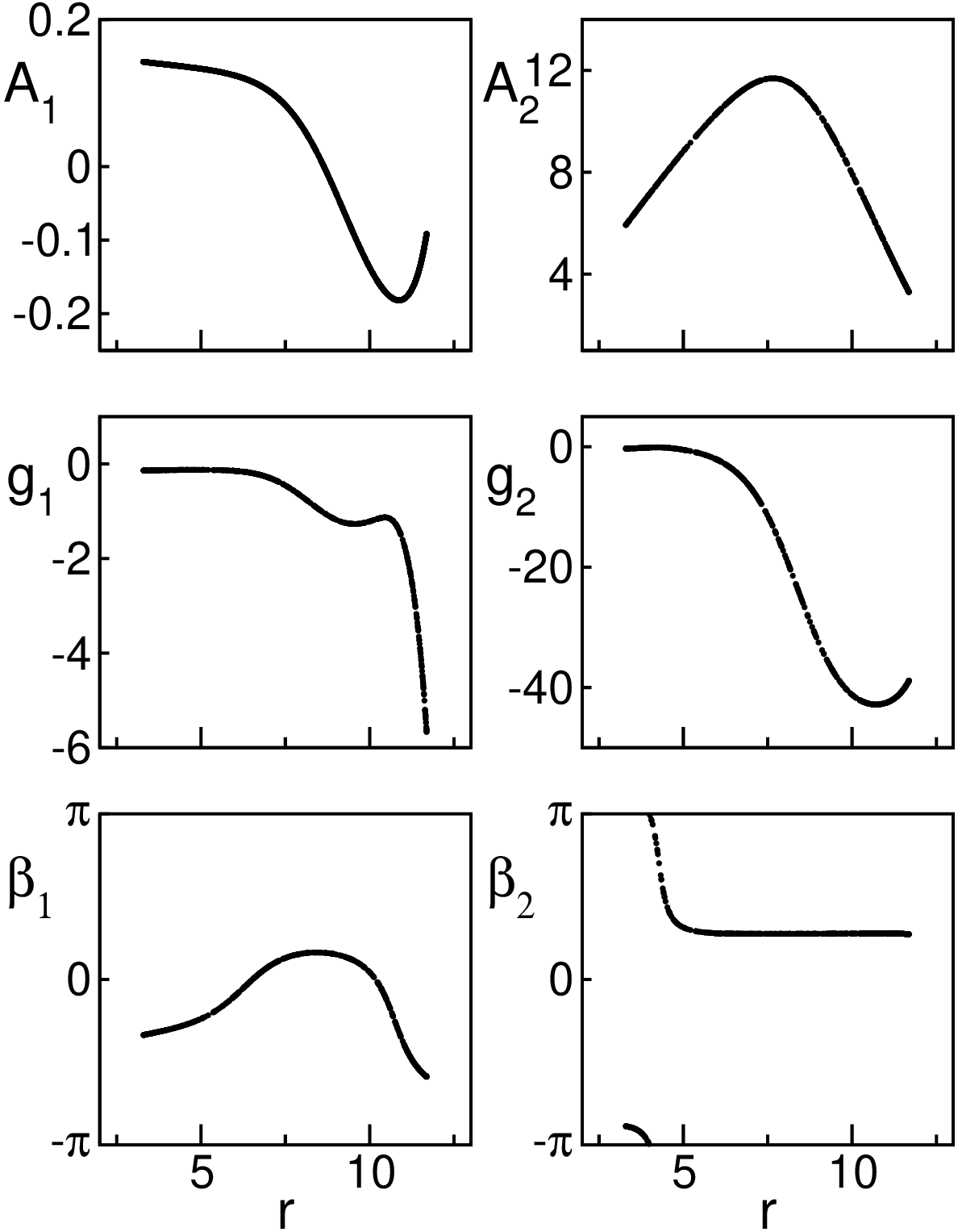}
\caption{Numerically computed functions $A_{i}$, $g_{i}$ and $\beta_{i}$
$(i=1,2)$ for the R\"ossler oscillator. R\"ossler parameters have been chosen
as in Ref. \cite{Pikov97b}: $a_0=0.2$, $a_1=1$ and $a_2=9$.  }
\label{cap:Numerically-computed-functions} 
\end{figure}

The correctness of the scheme is confirmed in
Fig.~\ref{cap:Numerically-computed-functions}, where all the functions defining
the model have been numerically obtained. The very fact that they all
look as one-dimensional curves, confirms the conjecture that $z$-dependence can
be neglected.

The GSF \toref{2dFlow2} generalizes the model introduced in Ref.~\cite{Pikov97b}, 
where the effect of the phase on the $r$ dynamics was not included. This implies
that the GSF looses the skew-product structure. This has important consequences
on the orientation of the second Lyapunov vector that we determine in
the next sections. Notice also that the GSF \toref{2dFlow2} generalizes and
justifies the model invoked in Ref.~\cite{Pikov97}.

In spite of the simplification introduced by removing the $z$ variable, a
rigorous treatment of Eq.~\toref{2dFlow2} for generic functions $g$ and $\beta$ is
still very difficult. A first obstacle may be the lack of a finite Markov
partition for the unperturbed system, which does not allow us expressing the
second order correction to the LV in a closed form (see appendix \ref{app2} for
details). A second obstacle is that the perturbation itself may and will in
general destroy the Markov partition, making the invariant measure hardly
accessible to a perturbative expansion. For both reasons, we restrict ourselves
to considering specific $A$, $g$, and $\beta$ functions which guarantee the
existence of a finite Markov partitions in a finite range of the coupling
constant. In the last section we shall comment on the possibility to extend
our formalism to a more general setup.

For the sake of simplicity, we have decided to analyse the following model,
\begin{eqnarray}
r' & = & f(r)+2\varepsilon cg(r)\cos(\psi+\alpha)\nonumber\\
\psi' & = & \psi+K +\Delta r+\varepsilon b\cos\psi
\label{GSF}
\end{eqnarray}
where
\begin{equation}
f(r)=1-2|r|\quad,\quad g(r)=r^{2}-|r|
\label{tentmap}
\end{equation}
with $r\in[-1,1]$. The tent-map choice for $r$ ensures that $[-1,0]$ and $[0,1]$
are the two atoms of a Markov partion. Moreover, since $g(r)$ is equal to 0 for
$r=0$ and $r=\pm 1$, this remains true also when the perturbation is switched
on. This is a key property that is necessary to perform a completely analytical
treatment in the following sections.

In this two-dimensional setup, the formal expression of the $i$th LE
\toref{genLyap1} writes
\begin{equation}
\lambda^{(i)}=\frac{1}{2} \int_{-1}^{1}dr\int_{0}^{2\pi}d\psi P(r,\psi) \ln\left[
\frac{||{\textbf{J}}(r,\psi){\textbf{V}}^{(i)}(r,\psi)||^{2}}
   {||{\textbf{V}}^{(i)}(r,\psi)||^{2}}\right]
\label{Lyap2}
\end{equation}
and the Jacobian is
\begin{equation}
{\textbf{J}}(r,\psi)=\left(\begin{matrix}f_{r}(r)+2\eps cg_{r}(r)\cos(\psi+\alpha) &
-2\eps cg(r)\sin(\psi+\alpha)\\
\Delta & 1-\eps b\sin\psi\end{matrix}\right)
\end{equation}
where the subscript $r$ denotes the derivative with respect to $r$.
The computation of the Lyapunov exponent therefore, requires determining both
the invariant measure $P(r,\psi)$ and the local direction of the
Lyapunov vector ${\textbf{V}}^{(i)}$. 

\section{A perturbative calculation of the second Lyapunov exponent}
\label{LL}
In this section we derive a perturbative expression for the second LE of the
GSF \toref{GSF}, by expanding Eq.~\toref{Lyap2}. One of the key ingredients is
the second LV, whose direction can be identified by writing
${\bf V} = (V,1)$ (for the sake of clarity, from now on, we omit the
superscript $i=2$ in $\bf V$ and $\lambda$, as we shall refer only to the second
direction). Due to the skew-product structure of the unperturbed map
\toref{GSF}, the second LV is, for $\varepsilon=0$, aligned along the $\psi$
direction (i.e. $V=0$). It is therefore natural to expand $V$ in powers of
$\eps$
\begin{equation}
V \approx \eps v_1(r,\psi) + \eps^2 v_2(r,\psi)
\label{Vexp}
\end{equation}

Accordingly, the logarithm of the norm of $\bf V$ is
\begin{equation}
\ln ||{\bf V}||^2 = \ln (1 + \eps^2 v_1^2) = \eps^2 v_1^2
\end{equation}
while its forward iterate writes as (including only those terms that contribute
up to second order in the norm),
\begin{equation}
{\bf J V} = \left( \begin{matrix}
\eps f_r(r) v_1 - 2 c \eps g(r) \sin(\psi+\alpha)  \\
        1 + \eps( \Delta v_1 - b \sin \psi) + \eps^2 \Delta v_2) 
\end{matrix} \right)
\label{eq:jacob1}                 
\end{equation}
Notice that we have omitted the $(r,\psi)$ dependence of $v_1$ and $v_2$ 
to keep the notation compact. 
 
The Euclidean norm of the forward iterate is
\begin{equation}
||{\bf JV}||^2 = 1 + 2\eps( \Delta v_1 - b \sin \psi) + \eps^2 \Big\{
 ( \Delta v_1 - b \sin \psi)^2 + 2 \Delta v_2 +
 [f_r(r) v_1 - 2 c g(r) \sin(\psi+\alpha)]^2 \Big\}
\end{equation}
and its logarithm is
\begin{equation}
\ln ||{\bf JV}||^2 = 2\eps( \Delta v_1 - b \sin \psi) - \eps^2 \Big\{ 
 (\Delta v_1 - b \sin \psi)^2 - 2 \Delta v_2 -
 [f_r(r) v_1 - 2 c  g(r) \sin(\psi+\alpha)]^2 \Big\}
\end{equation}
We now proceed by formally expanding the invariant measure in powers of $\eps$
\beq
P(r, \psi) \approx p_0(\psi) + \eps p_1(r, \psi) + \eps^2 p_2(r, \psi)  .
\label{Pexp}
\eeq 
The determination of the $p_i$ coefficients is presented in the next section,
but here we anticipate that, as a consequence of the skew-product structure
for $\varepsilon=0$, the zeroth-order component of the invariant measure
does not depend on the phase $\psi$. Moreover, because of the structure of the
tent-map, $p_0$ is also independent of $r$, i.e. $p_0 = 1/ 4 \pi$.
The second Lyapunov exponent can thus be written as
\begin{eqnarray}
\label{Lyap3}
\lambda &=& \int_{-1}^1 dr \int_0^{2\pi} d \psi
\Big( \frac{1}{4 \pi} + \eps p_1(r, \psi) \Big) \Big\{
2\eps( \Delta v_1(r, \psi) - b \sin \psi) - \eps^2 \Big[(\Delta v_1(r, \psi) - 
b \sin \psi)^2 \nonumber \\
&&  - 2 \Delta v_2(r, \psi) +
 [f_r(r) v_1(r, \psi) + 2 c g(r) \sin(\psi+\alpha)]^2 + v_1^2(r,\psi)\Big] \Big\}
  + o(\eps^2)
\end{eqnarray}
As the variable $\psi$ is a phase, it is not a surprise that some
simplifications can be found by expanding the relevant functions into Fourier
components. We start writing the first component of the invariant measure as
\beq
p_1(r, \psi) = \frac{1}{2\pi} \sum_{n} q_i(r) e^{i n \psi}
\label{qFourier}
\eeq
We then turn our attention to the first order component $v_1(r,\psi)$
of the second LV \toref{Vexp}. Due to the sinuosidal character of the forcing
term in the GSF \toref{GSF}, it is easy to verify (see the next section) that
$v_1(r,\psi)$ contains just the first Fourier component,
\begin{equation}
v_1(r, \psi) = c\Big[ L(r) \sin(\psi+\alpha)  + R(r) \cos(\psi+\alpha)\Big]
\label{eq:ansatz}
\end{equation}
By now, inserting Eqs.~(\ref{qFourier},\ref{eq:ansatz}) into Eq.~\toref{Lyap3} and
performing the integration over $\psi$, we obtain

\begin{eqnarray}
\label{Lyap4}
\lambda &=& \eps^2 \int_{-1}^1 dr \left\{ \Delta c \Bigg[ q_1^r \Big[ L(r) \sin \alpha +
R(r) \cos \alpha \Big]
- q_1^i \Big[ L(r) \cos \alpha - R(r) \sin \alpha \Big] \Bigg] \right. \nonumber  \\
&&+ b q_1^i  - \frac{b^2}{8}
 + \Delta \frac{bc}{4} \Big[ L(r) \cos \alpha - R(r) \sin \alpha \Big]
+ \frac{c^2}{8}(3-\Delta^2) \Big[L^2(r)+R^2(r)\Big]
\label{eq:lambda} \\
&&\Bigg. + \frac {c^2}{2} g^2(r)
 + c^2 \frac{|r|}{r} g(r)L(r)
 \Bigg\}  + \frac{\Delta I_2}{4\pi} \nonumber
\end{eqnarray}
where we have further decomposed $q_1(r)$ in its real and imaginary parts
\beq
q_1(r) = q^r_1(r) + i q^i_1(r)
\eeq
and we have defined
\begin{equation}
I_2 := \int_{-1}^1 dr \int_0^{2\pi} d\psi \, v_2(r,\psi),
\label{eq:i2}
\end{equation}
which accounts for the contribution arising from the second order correction
to the LV. This expansion shows that the highest-order contribution to the
second Lyapunov exponent of the GSF scales quadratically with the perturbation
amplitude. This is indeed a general result that does not depend on the
particular choice of the functions used to define the GSF, but only on the
skew-product structure of the unperturbed time evolution and on the validity
of the expansion assumed in \toref{Pexp} (we shall comment on this last issue
in the next section).

By inserting the expression for $I_2$ obtained in appendix \ref{app2} (see
Eq.~\toref{eq:stot2}) in Eq.~\toref{Lyap4}, we finally obtain the perturbative
expression for the second LE,
\begin{eqnarray}
\lambda = \eps^2 \Bigg\{ \frac{c^2}{30}-\frac{b^2}{4} + \int_{-1}^1 dr \Bigg[ b q_1^i(r) +
 \frac{c^2}{16}(6-\Delta^2) \Big[L^2(r)+R^2(r)\Big] \Bigg. \Bigg. \label{eq:lambda:fin}\\
+ \Delta c q_1^r(r) \Big[ L(r) \sin \alpha + R(r) \cos \alpha \Big]
+ \Delta c \left(\frac{b}{4} - q_1^i(r)\right) \Big[ L(r) \cos \alpha - R(r)
\sin \alpha \Big] \nonumber  \\
\Bigg. \Bigg. + c^2 \frac{|r|}{r} g(r)L(r) + \frac{\Delta c^2}{4} r
\sin \left( \frac{\Delta(1-r)}{2} \right) \Big[ L(r) \cos K  - R(r)
\sin K \Big] \Bigg]\Bigg\}
\nonumber
\end{eqnarray}
Accordingly, the numerical value of the second LE can be obtained by performing
integrals which involve the four functions $q^r_1(r)$, $q^i_1(r)$, 
$L(r)$, and $R(r)$,
that are determined in the next section.

\section{Determining the coefficients of the power expansion}
\label{measure_lyap}
After having more or less formally expanded the expression of the second LE in
powers of the coupling strength $\varepsilon$ in the previous section, now
we show how the coefficients can be determined for both the invariant measure
and the direction of the LV. Notice that the second part of the project passes
through the implementation of the general ideas put forward in Sec. II.

\subsection{The invariant measure}
We start focusing our attention on the invariant measure $P(r,\psi)$ which can
be computed as a fixed point of the Frobenius-Perron equation 
\begin{equation}
\label{fpeq}
P'(r',\psi') = \frac{P(r^-,\psi^-)}{|\det {\bf J}(r^-,\psi^-)|} +
\frac{P(r^+,\psi^+)}{|\det {\bf J}(r^+,\psi^+)|} 
\end{equation}
where $(r^-,\psi^-)$ and $(r^+,\psi^+)$ are the two preimages of
$(r',\psi')$. It is important to notice that our choice of the map guarantees
that two solutions do exist in the whole rectangle $[-1,1]\times[0,2\pi]$ in a
finite range of $\varepsilon$-values. This will be crucial for obtained exact
expressions. As it has been shown in the previous section, we are interested in
solving the above equation up to first order. Accordingly, we write
\begin{equation}
\label{expprobeq}
P(r',\psi') = p_0(r',\psi') + \frac{\eps}{2\pi} \sum_{n} q_i(r) e^{i n \psi} 
\end{equation}
where we have expanded the first order contribution as in Eq.~\toref{qFourier}. 
It is also necessary to expand the preimages
\begin{eqnarray}
r^\pm &=& r^\pm _0 + \varepsilon r^\pm_1 \\
\psi^\pm &=& \psi^\pm _0 + \varepsilon \psi^\pm_1
\end{eqnarray}
where
\begin{eqnarray}
r^\pm_0 &=&  \pm \frac{1-r'}{2}\\
\psi^\pm_0 &=&  \psi' - K \mp \Delta \frac{1-r'}{2}\\
r^\pm_1 &=&  \mp c\frac{1-r'^2}{4}\cos(\psi^\pm_0+\alpha) \\
\psi^\pm_1 &=&  \pm c\Delta \frac{1-r'^2}{4}\cos(\psi^\pm_0+\alpha)
  -b \cos{\psi^\pm_0}
\end{eqnarray}
At zeroth order in $\varepsilon$, it is easy to see that the Frobenius-Perron
equation (\ref{fpeq}) reduces to
\begin{equation}
p_0(r',\psi') = \frac{1}{2} \big( p_0(r^+_0,\psi^+_0) + p_0(r^-_0,\psi^-_0) 
 \big)
\end{equation}
whose solution is everywhere constant, as anticipated in section \ref{LL}. 
By imposing the normalization condition, one obtains
\begin{equation}
p_0 = \frac{1}{4\pi} .
\end{equation}
By then considering that
\begin{equation}
|\det {\bf J}(r,\psi)|^{-1} = \frac{1}{2} \big[ 1 +  \varepsilon
  b \sin \psi + \varepsilon c \, {\rm sign}{r} ( \Delta g(r) \sin(\psi + \alpha) +
  g_r(r) \cos(\psi + \alpha))\big]
\end{equation}
and projecting Eq.~(\ref{fpeq}) over its first Fourier component, we finally
obtain a closed equation for $q_1(r)$
\begin{eqnarray}
q_1(r') = \frac{{\rm e}^{-iK}}{2} \left[ 
{\rm e}^{-i\Delta r^+_0} 
  \left( q_1(r^+_0)- i\frac{b}{4} + \frac{c}{4}
{\rm e}^{i\alpha} g_r(r^+_0) - i \frac{c}{4}{\rm e}^{i\alpha} \Delta
  g(r^+_0) \right) + \right. \\ 
\left. {\rm e}^{-i\Delta r^-_0} 
  \left( q_1(r^-_0)- i\frac{b}{4} - \frac{c}{4}
{\rm e}^{i\alpha} g_r(r^-_0) + i \frac{c}{4}{\rm e}^{i\alpha} \Delta
  g(r^-_0) \right) \right]
\end{eqnarray}
The structure of this equation is very similar to a Frobenius-Perron equation
for a one-dimensional system. The dimensionality reduction has been made
possible by exploting the skew-product structure of the unperturbed system.
Considering also the simple expression of the preimages of $r'$ (they
have to be determined at zeroth order), the above equation can be accurately
solved by implementing the standard method to solve a Frobenius-Perron equation
(the only limit being imposed by the numerical round-off).

\begin{figure}
\includegraphics*[width=9.cm]{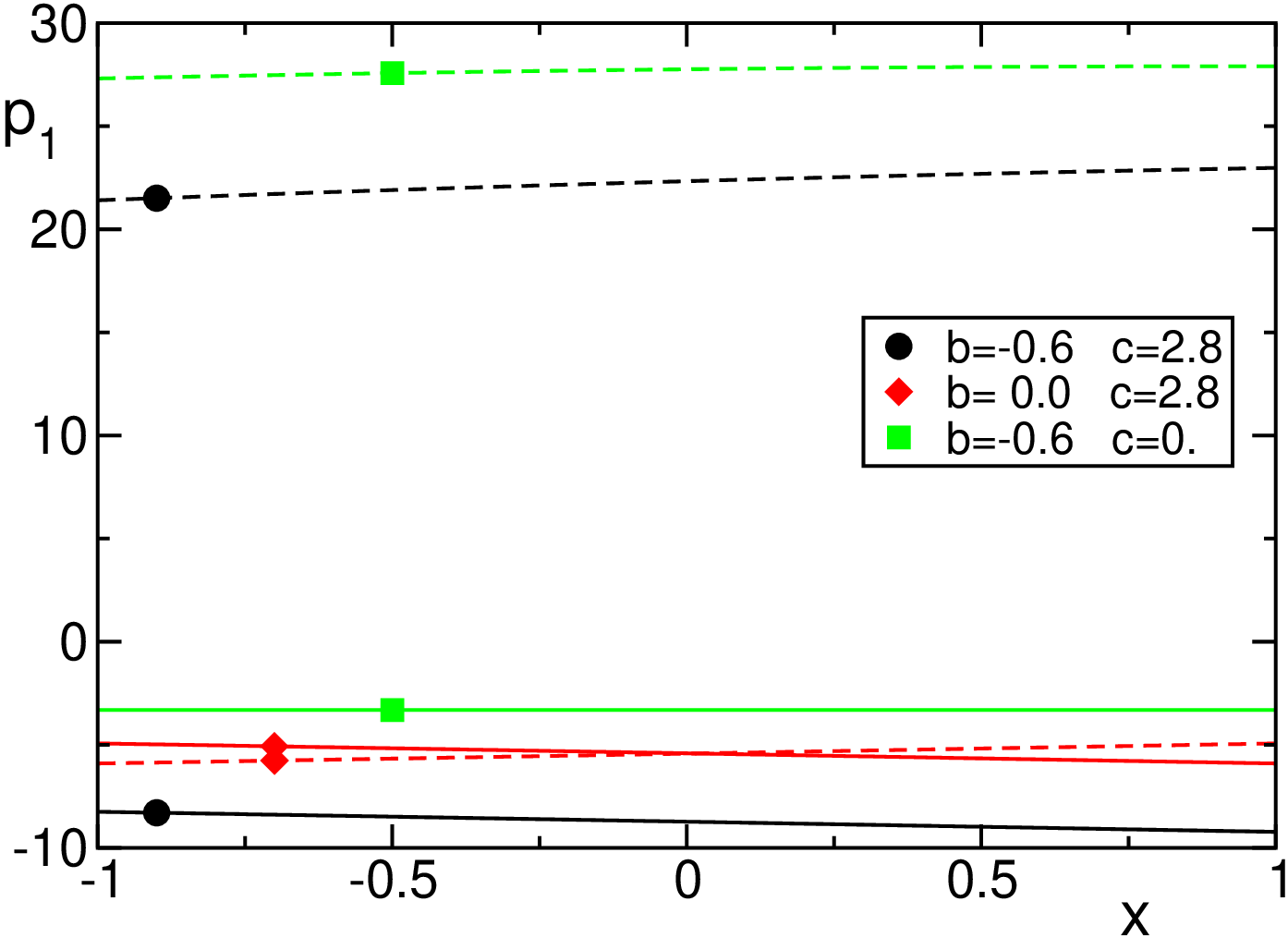}
\caption{The first order contribution $q_1$ to the probability density for the
parameter values: $\Delta = -0.18$, $\alpha=-\pi/4$, $K = 0.04 \pi$. Solid and
dashed lines refer to real and imaginary parts.}
\label{fig:p1}
\end{figure}

In Fig.~\ref{fig:p1} we have plotted the real and immaginary parts of $q_1$ 
for three different choices of the parameters $b$ and $c$. In all cases, one can
see a very smooth dependence, which thus suggests the possibility to obtain
accurate fully analytic expressions by expanding polynomially $q_1(r)$.
However, being more interested in testing the overall validity of the
perturbative approach, we do not explore this possibility.

In fact, in order to test the general validity of the power expansion, we
have numerically investigated three different GSFs, corresposinding to the
following choices of the functions $f$ and $g$: {\it i)} $f(r)=1-2|r|$, 
$g(r)=r^{2}-|r|$ as considered in \toref{tentmap}; {\it ii)} $f(r)=0.8-1.8|r|$ and
$g(r)=1/2$, for which there is no finite Markov partition; {\it iii)}
$f(r)=2(1-2 \eps c) (1-|r|)$ and $g(r)=1/2$, for which the finite Markov
partition existing in the unperturbed case is destroyed as soon as the
perturbation is switched on.

\begin{figure}
\includegraphics*[width=9.cm]{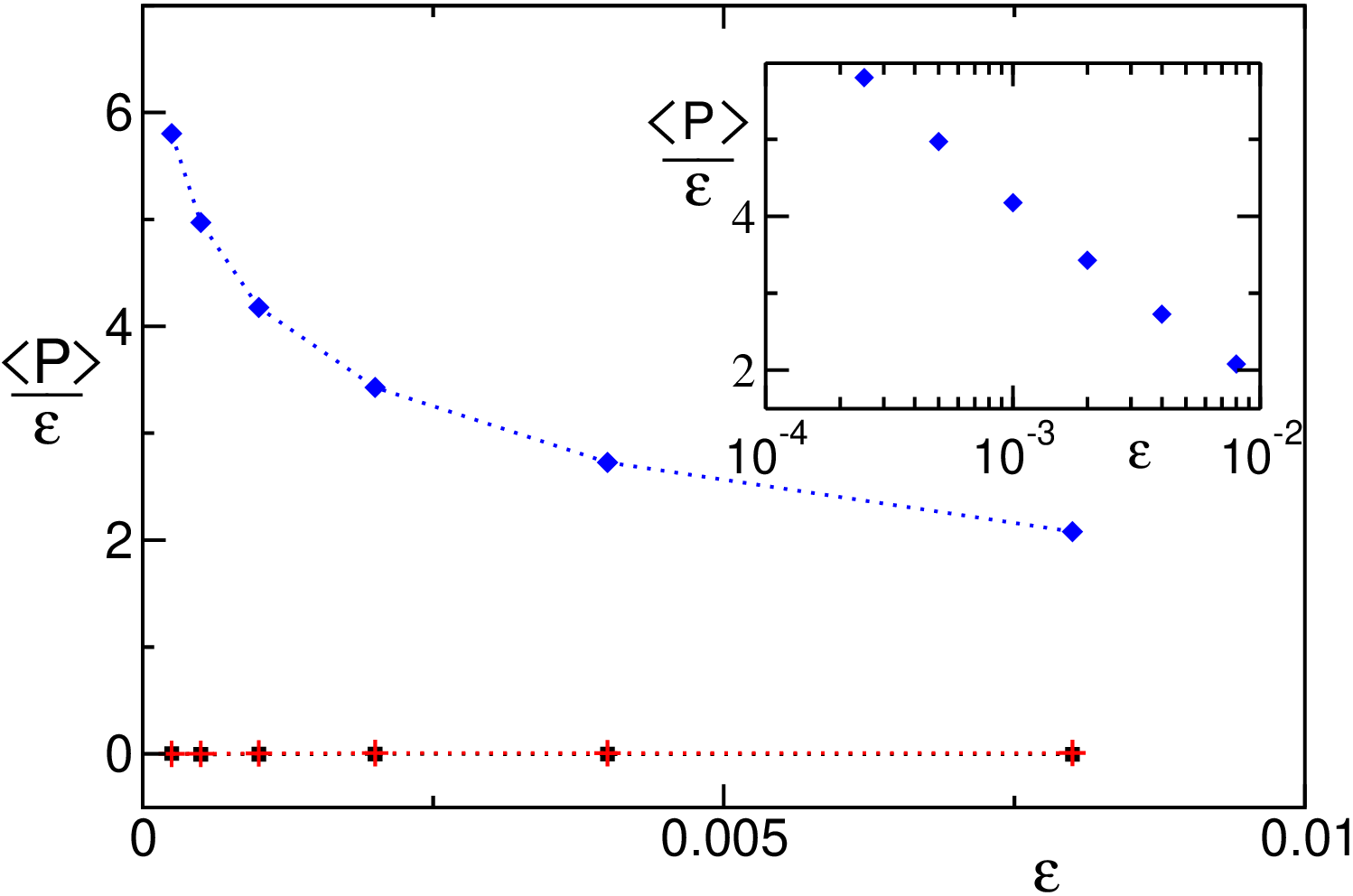}
\caption{The deviation of the zero Fourier component of the invariant measure
(see Eq.\toref{pippo} of map \toref{GSF} as a function of $\varepsilon$. Three
different choices of $f$ and $g$ have been tested: i) $f(r)=1-2|r|$ and
$g(r)=r^{2}-|r|$ (crosses), ii) $f(r)=0.8-1.8|r|$ and $g(r)=1/2$ (squares), and
iii) $f(r)=2(1-2 \eps c) (1-|r|)$ and $g(r)=1/2$ (diamonds). 
Parameter values have been fixed to $\Delta=-0.18$, $\alpha=-\pi/4$,
$K=0.04\pi$, $b=-0.6$ and c=$2.8$. Abscissas are
divided by $\eps$ to better emphasize deviations from linear scaling. The
logarithmic deviations dispalyed by the diamonds are emphasized in the inset.}
\label{MMap}
\end{figure}

In order to compare such models, we have computed the deviation of the
zero Fourier component of the invariant measure of the map \toref{GSF}
induced by a small finite coupling $\varepsilon$, 
\beq
\langle P(r,\psi) \rangle = \int_{-1}^1 dr \int_0^{2\pi} d \psi 
\Big[P(r,\psi)|_{\eps} - P(r,\psi)|_{\eps=0}\Big]  .
\label{pippo}
\eeq
As it can be clearly seen in Fig.~\ref{MMap}, in the first two cases the linear
term is even equal to zero, while relevant multiplicative logarithmic
corrections are present in the third case. This ``pathological" behaviour is
induced by the fact that as soon as the coupling is switched on, an infinite
series of discontinuities in the invariant measure suddenly arises 
in the vicinity of the former fixed point $r=-1$. It is, however, important to
notice that no peculiarity is found in the more generic second case.

\subsection{The direction of the second Lyapunov vector}
\begin{figure}
\includegraphics*[width=9.cm]{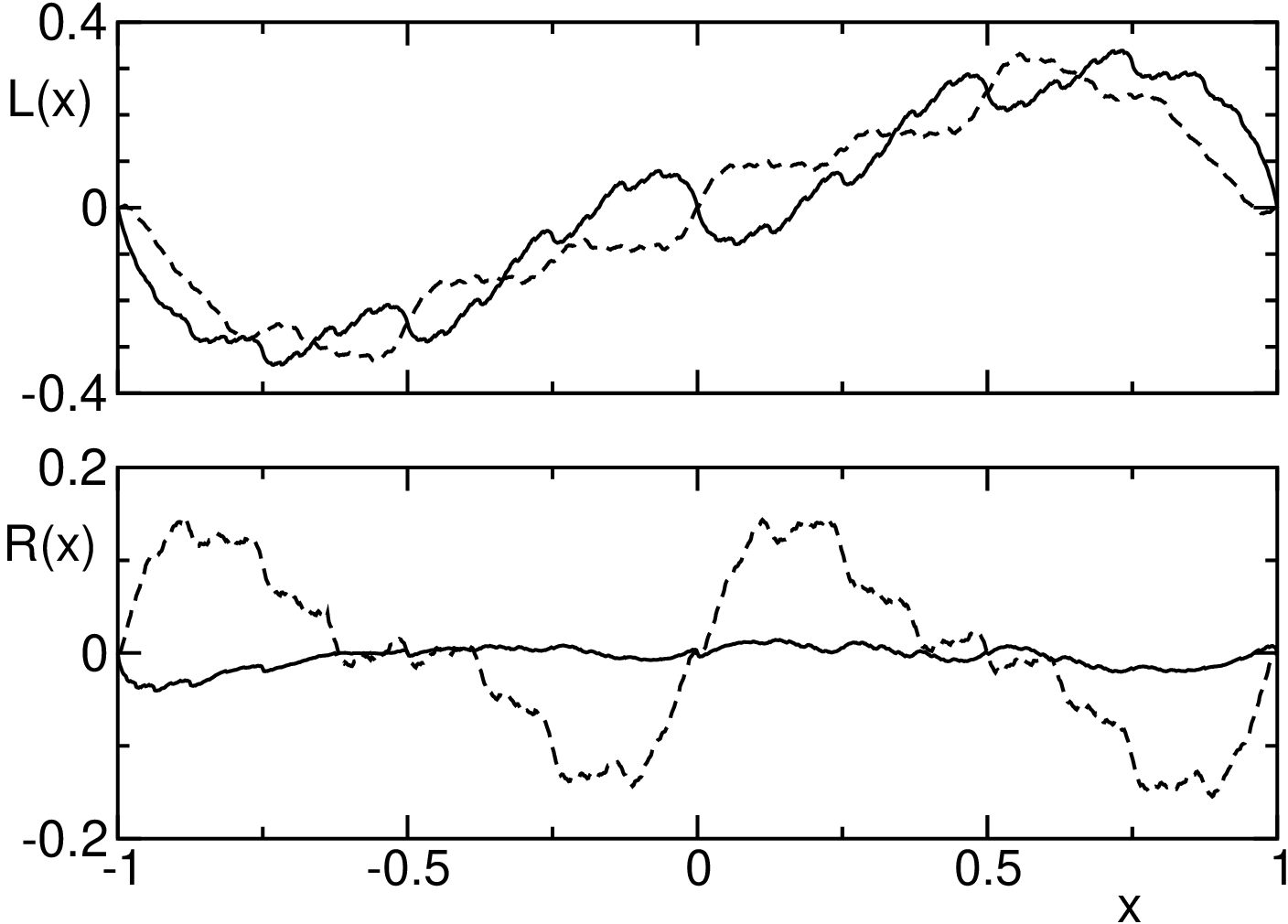}
\caption{The functions $L(r)$ and $R(r)$ computed for $\Delta = -0.18$ and
$K=0$ (solid curve) and $K=-1$ (dashed line).}
\label{fig:lr}
\end{figure}

In this subsection we derive a self-consistent equation for the second LV.
We start from Eq.~(\ref{eq:jacob1}), retaining only the relevant perturbation
terms 
\begin{equation}
{\bf J V} = \left( \begin{matrix}
  [f_r(r) + 2 \eps c  g_r(r) \cos(\psi+\alpha)]V -2 \eps c g(r)
                                                          \sin(\psi+\alpha)  \\
        \Delta V + 1 - \eps b \sin \psi
\end{matrix} \right)
\end{equation}
By computing the ratio between the components of ${\bf J V}$ we obtain the
new value of the slope $V'=\eps v'_1 + \eps v'_2 + \ldots$,
\begin{equation}
v'_1 + \eps v'_2 =
\frac{\Big[f_r(r) + 2\eps c g_r(r)\cos(\psi+\alpha)\Big] (v_1 + \eps v_2)
                -2 c g(r)\sin(\psi+\alpha)}
        {\Delta \eps v_1 + (1 - \eps b \sin \psi)},
\label{eq:vec1}
\end{equation}
where we have again kept only the relevant terms (up to first order after
dividing both sides by $\eps$) and
where $v'_1$ and $v'_2$ are both functions of the iterates $r'$ and $\psi'$,
\begin{eqnarray}
r' = r'_0 + 2\eps c g(r) \cos(\psi + \alpha) \\
\psi' = \psi'_0 + \eps b \cos \psi
\end{eqnarray}
where $r'_0 = f(r)$ and $\psi'_0 = \psi + K + \Delta r$
are the unperturbed iterates.
By replacing the expressions for $r'$ and $\psi'$ in Eq.~(\ref{eq:vec1}), at
leading order, we obtain

\begin{equation}
v'_1(f(r),\psi +K+\Delta r) = f_r(r) v_1(r,\psi) - 2cg(r) \sin(\psi+\alpha)
\label{eq:rec}
\end{equation}
As anticipated in Sec.~\ref{sec2}, this recursive relation can be solved by
following backwards the dynamics of $(r,\psi)$. It is worth stressing that
the term $2cg(r) \sin(\psi+\alpha)$, i.e. the effect of the phase on the
amplitude dynamics, acts as a source term in Eq.~\toref{eq:rec}. In its absence,
the latter equation would yield a trivial vanishing solution for $v_1$ (which in
turn also implies $v_2=0$, as it can be appreciated in App.~\ref{app2}).
It is therefore the feedback of the phase on the amplitude dynamics that
generates a nontrivial orientation of the perturbed second Lyapunov vector.
Furthermore, the structure of the source term $2cg(r) \sin(\psi+\alpha)$
naturally suggests the Ansatz (\ref{eq:ansatz}).
By inserting it in Eq. \toref{eq:rec} we obtain two recursive equations,
\begin{eqnarray}
L(r) = {\rm sign}(r) \left[ \frac{1}{2} \Big (R(f(r)) \sin(K+\Delta r) -
  L(f(r)) \cos(K+\Delta r)\Big) - g(r) \right] \nonumber \\
R(r) = -\frac{1}{2} {\rm sign}(r) \left[(R(f(r)) \cos (K+\Delta r) +
  L(f(r)) \sin(K+\Delta r) \right]
\end{eqnarray}
This equation can be solved numerically, by considering it as a recursive
relation to be iterated backward in time until the fixed point solution is
eventually attained and the functions $L$ and $R$, computed with the desired
precision. In Figs.~\ref{fig:lr} we can see some examples of how they look like.

>From the analysis carried on in Sec.~\ref{sec2} and in particular from
Eq.~\toref{bareta}, we see that the condition for a smooth behaviour of the
direction $V$ along the phase-direction is (noticing that here,
$\gamma = \lambda_2-\lambda_1$) is $\lambda_1>2\lambda_2$ that is certainly
verified and this fully justifies the expansion in Fourier modes along such a
direction. On the other hand, along the expanding direction $r$, the codition
writes $\lambda_2<0$, that is only marginally verified. The apparent roughness
exhibited by $L(r)$ and $R(r)$ can therefore be a manifestation of the expected
non-complete smoothness. It is, however, also important to stress the role
played by the functions we have specifically considered in the GSF.
In fact, the tent map induces a discontinuity in the tangent space that
propagates everywhere, though significantly squeezed. Luckily enough, as it can
be appreciated in App.~\ref{app2}, such singularities are integrated out when
determining the leading contribution to the Lyapunov exponent which is therefore
substantially insensitive and can be computed without much harm.

\section{Numerical results and conclusions}
\label{conc}

In this paper we have introduced a novel approach to determine analytically
the Lyapunov exponent and applied it to the specific case of a periodically
forced chaotic oscillator described by a model (the generalized special flow)
which is also introduced here starting from the specific case of the R\"ossler
oscillator.

\begin{figure}
\includegraphics*[width=9.cm]{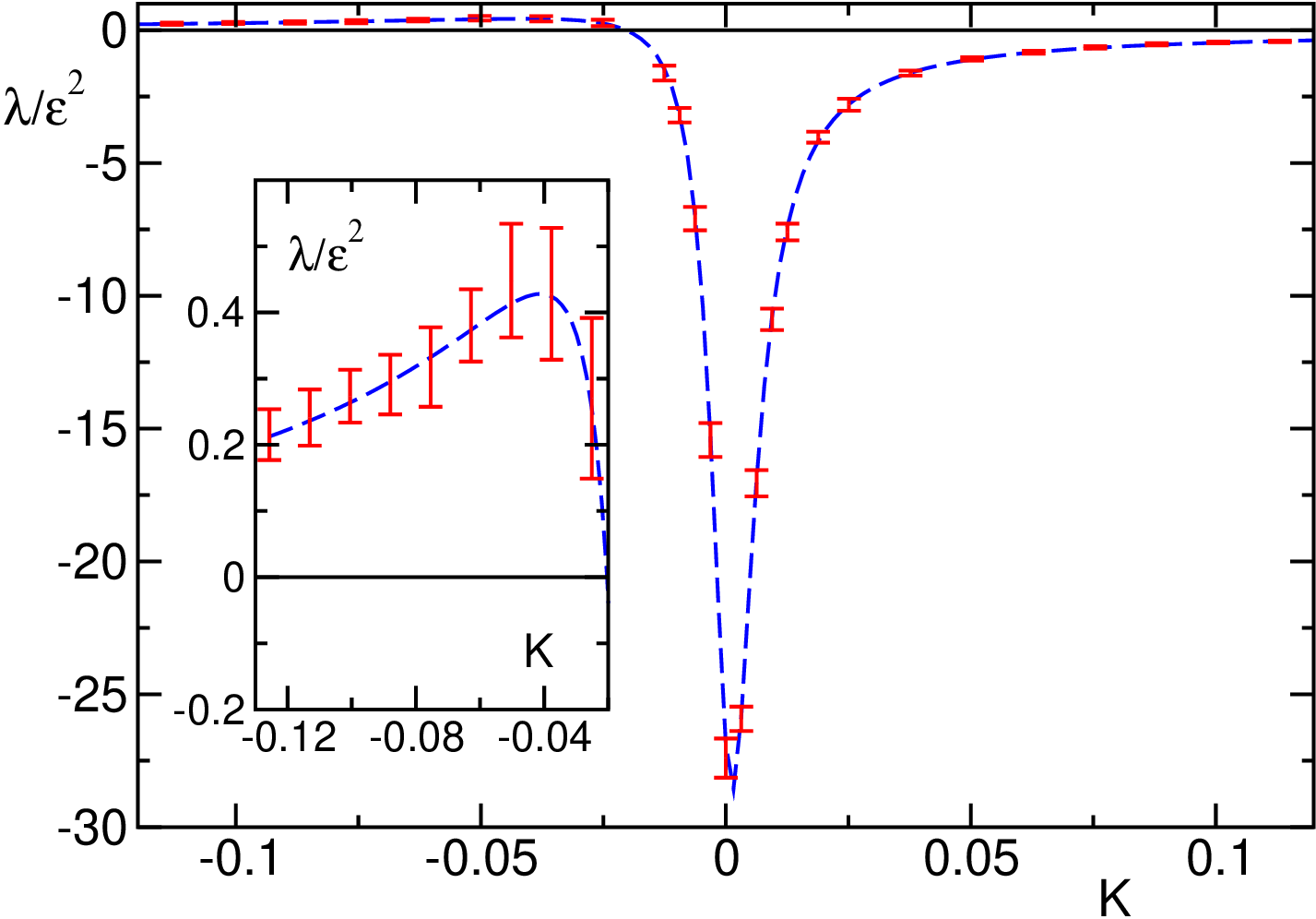}
\caption{Second Lyapunov exponent for the Generalized Special Flow \toref{GSF}
as a  function of detuning $K$. The dashed line represent the analytical result,
while dots (with the bars indicating one standard deviation) results of direct
numerical simulations of the GSF. Parameter values have been fixed to:
$\Delta=-0.18$, $\alpha=-\pi/4$, $b=-0.6$, $c=2.8$. Abscissas have been rescaled
by a factor $\eps^2$ to evidentiate the second order coefficient. The inset
magnifies a part of the larger graph.}
\label{fig:scal1}
\end{figure}

Given the many technical difficulties that is necessary to overcome in order to
finally obtain the numerical value of the quadratic coefficient, it is wise to
compare the analytic expression with the direct numerical computation of the
second LE performed for small enough values of $\eps$. In Fig.~\ref{fig:scal1},
the second order coefficient $\lambda/\varepsilon^{2}$ is determined from the
analytic expression \toref{eq:lambda:fin} and by directly simulating the GSF
for $\eps$-values in the range $[10^{-4},10^{-2}]$. The good agreement obtained
for all $K$ values confirm
the correctness of the analytical calculations. The relative strong negative
peak around $K=0$ is a manifestation of a resonance phenomenon. The LE tends
to be more negative when the forcing frequency is close to the average frequency
of the chaotic attractor.

It is also important to stress that our results are valid for arbitrarily small
$\eps$, i.e. below the transition to phase-synchronization (if there is any)
and therefore tells us that the LE corresponding to the phase dynamics is
immediately different from zero, as soon as the coupling is switched on.

Another important point concerns the sign of the LE: naive considerations might
suggest that the coupling tends to stabilize the phase dynamics and thereby 
to give a negative LE. However, the left tail in Fig.~\ref{fig:scal1} (see also
the inset) definitely shows a positive exponent. It is desirable to find some
simple heuristic arguments to understand when and why the phase dynamics is
stable, but this does not seem to be an easy task and is left as an open problem
for future investigations.

\begin{figure}
\includegraphics*[width=9.cm]{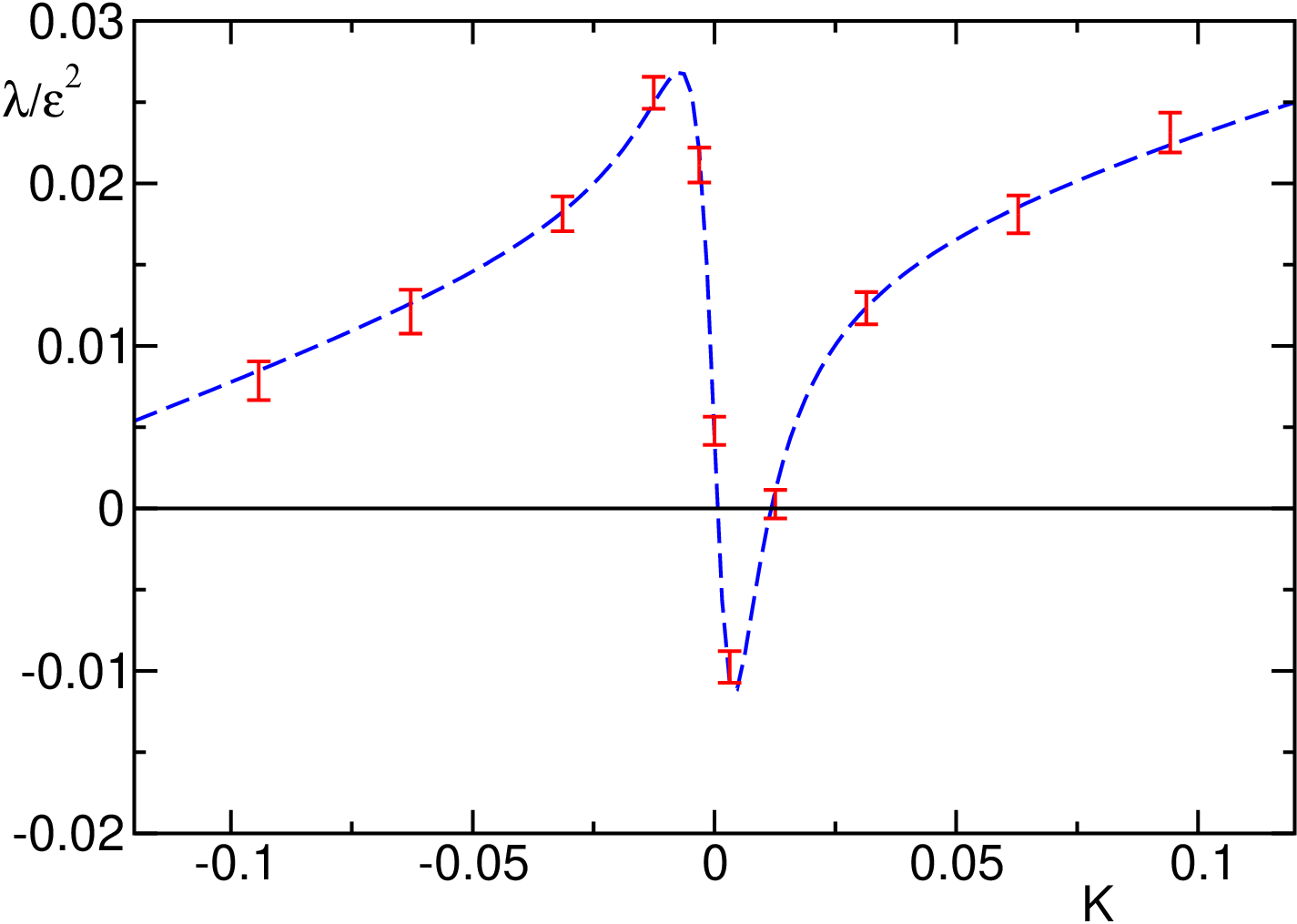}
\caption{Second Lyapunov exponent for the Generalized Special Flow \toref{GSF}
as a  function of detuning $K$. Symbols and parameters values are the same as in
Fig. \ref{fig:scal1}, except for $b=0$.}
\label{fig:scal2} 
\end{figure}

The major difference between the GSF, we analyse in this manuscript and the
special flow introduced in \cite{Pikov97b} is the term proportional to $c$ in the
equation for $r$ in Eq.~\toref{GSF}. Such a term prevents the possibility of
further dimension reductions and requires setting up the machinery we have
developed in this paper. It is therefore interesting to quantify its direct
effect on the actual value of the LE. This can be simply done, by setting the
other coupling term $b=0$, an assignment that is basically complementary to what
done in the standard special flow. The results, reported in Fig.~\ref{fig:scal2}
show a sort of ``dispersive" behaviour for the LE which also tends to be
positive. This suggests that the back coupling of the phase dynamics onto
the amplitude evolution maybe responsible for an eventually positive LE.

While Eq.~\toref{eq:lambda:fin} cannot by any means capture the quantitative
behavior of the original R\"ossler system, still the quadratic behaviour of the
second LE seems to be a very general feature even though we can imagine that
the lack of structural stability of generic oscillators may mask the overall
behaviour with the presence of additional stability windows. We have therefore
computed directly the second Lyapunov exponent for the periodically forced
R\"ossler system chosing the same set of parameters 
($a_0=0.2$, $a_1=1$ and $a_2=9$) considered in Ref. \cite{Pikov97b}.

\begin{figure}
\includegraphics[scale=0.5,angle=0]{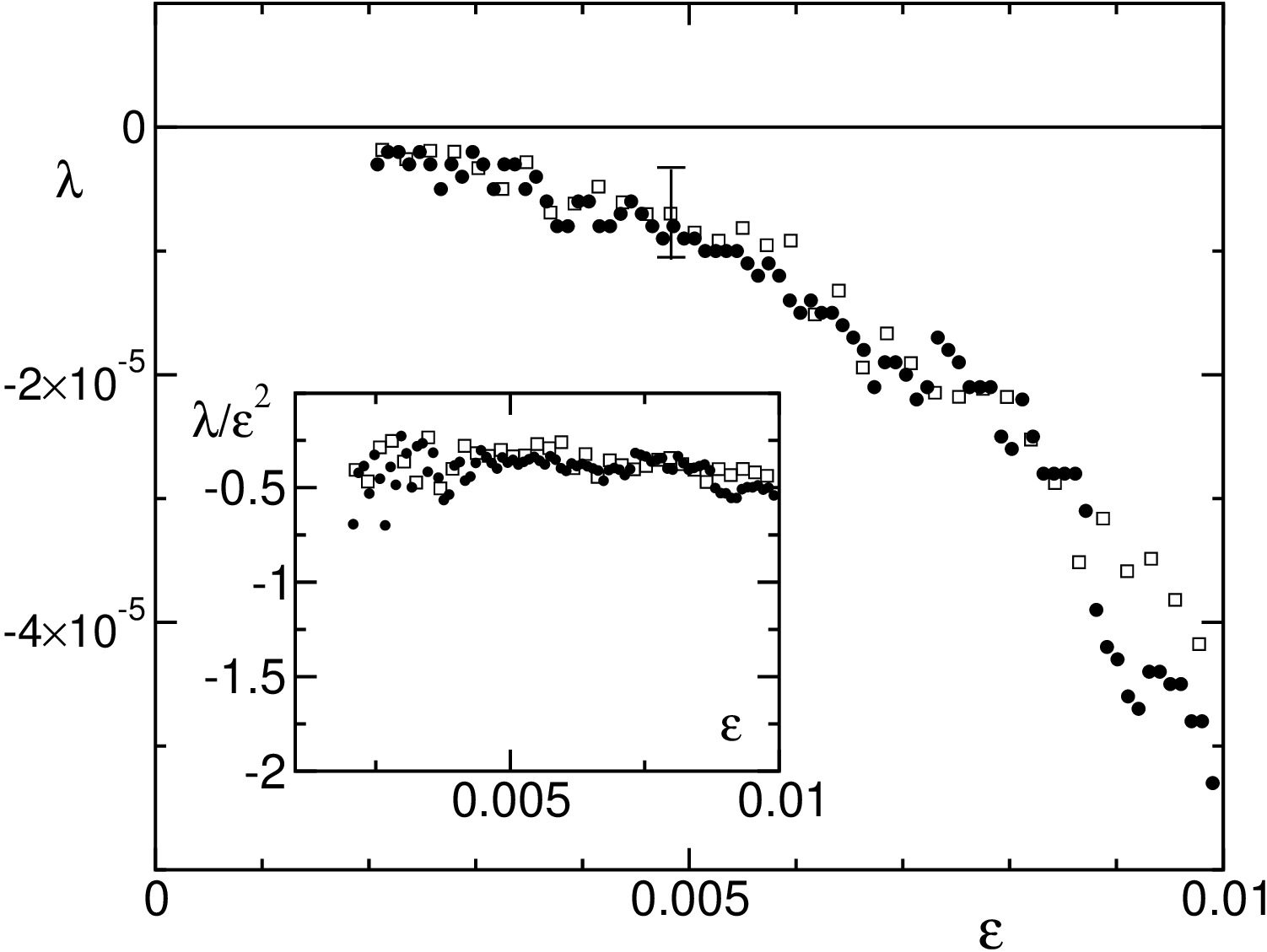}
\caption{Second Lyapunov exponent for the periodically forced R\"ossler system
close to the resonance. Full circles refer to a forcing frequency 
$\Omega=1.007$ (the R\"ossler natural average frequency is
$\nu_0=1.0158(1)$), while open squares correspond to 
$\Omega=1$. R\"ossler parameters have been fixed as $a_0=0.2$, $a_1=1$,
$a_2=9$, while  the integration interval is about $t=10^{8}$.  The inset shows
the quadratic relation between $\lambda^{(2)}$ and $\eps$.  }
\label{cap:rossl1}
\end{figure}

When the frequency of the periodic force is close to the natural frequency of
the oscillator, $\nu_0 =  1.0158(1)$ for our choice of parameters, 
we are able to detect the quadratic behavior
with a good accuracy, as demonstrated in Fig. \ref{cap:rossl1}. Since the
coupling strengths we have reached are much below the onset of 
phase synchronization, 
as can bee seen in Ref. \cite{Pikov97b}, these
numerical results confirm our theoretical conclusions that the Lyapunov exponent
corresponding to the phase dynamics deviates from zero as soon as the coupling
is switched on. It would be now interesting to extend the analysis carried out
in this paper to two coupled chaotic oscillators, perhaps by investigating
suitable discrete-time models such as that one introduced in \cite{FUH}. However,
while one can presumably learn something on the phenomenon of
phase-synchronization, we do not expect qualitative changes 
for the behaviour of the Lyapunov spectrum. This is supported by the recently
reported quadratic growth of the fifth LE (the first one corresponding to phase
dynamics) in a system of four coupled R\"ossler oscillators \cite{Popovych05}.

Altogether, the numerical and
analytical results presented in this paper clearly show that the onset of
phase-sinchronization is not signalled by the second LE (or, more generally,
the LE associated to the phase dynamics) turning negative, but it is rather
associated to a change in the structure of the dynamical attractor
\cite{Pikov97, Ott98} that is not directly related to the sign of the
``phase exponent". On the other hand, the quadratic dependence on the coupling
strength makes it difficult to numerically appreciate deviations from zero
(especially because of the statistical fluctuations that necessarily affect
numerical simulations) and explains why in earlier studies, the LE has been
mistakenly regarded as a proper order parameter to characterize the transition
to a phase-synchronized regime. 

Another important point concerns the sign of the second Lyapunov exponent. In
fact, it was formerly believed that phase chaos (i.e. a positive LE) can only
occur in the presence of a specific structure of the underlying chaotic
attractor (see e.g. \cite{Lai03}). On the other hand, our analytical results show
that the second LE can be positive even in a context where no peculiar
amplitude evolution has to be invoked. However, our approach does not give any
physical insight about the expected sign of the LE. It will be certainly useful
to find under which conditions a chaotic phase dynamics may arise.

Finally another major achievement of this paper is that Lyapunov exponents can
be effectively determined from ensemble averages, passing through the
determination of the local direction of the corresponding Lyapunov vectors. 
From a purely numerical point of view, there is no conceptual difficulty to
applying this method for a more detailed characterization of high-dimensional
chaos \cite{Chate}. However, in the perspective of obtaining more general
analytical results, it is desirable to go beyond systems with finite Markov
partitions.

\acknowledgments
Part of this project has been carried out with the help of financial support
provided by the Collaborative Research NATO grant PST.CLG.979410.

\appendix
\section{From continuous to discrete time}
\label{app1}

In this appendix we present the detailed calculations relative to the
determination of the Poincar\'e mapping \toref{3dFlow} for the periodically
forced R\"ossler oscillator \toref{rossler}. Notice, however, that the
methodology is quite general and is indeed applicable to a generic periodically
perturbed system, as long as it can be written in the form (\ref{eqtot}).

Let us start by introducing some useful notations
\begin{equation}
\mathbf{U}(r,z,\psi;t)\equiv(\Phi(r,z,\psi;t),R(r,z,\psi;t),Z(r,z,\psi;t))
\end{equation}
denotes the phase point in $\mathbf{S_{c}}$ at time $t$ of a trajectory
started in $(0,r,z)$ at time 0 and with an initial phase of the forcing
term equal to $\psi$ (pay attention to the fact that the triple
$(r,z,\psi)\in\mathbf{S_{d}}$). The crossing time with the Poincar\'{e} surface
is determined by imposing that the phase $\Phi$ has increased by $2\pi$, i.e.,
\begin{equation}
\Phi(r,z,\psi;T)=2\pi.
\label{defT}\end{equation}
As we are interested in the small coupling regime, we can expand $\textbf{U}$
in powers of $\varepsilon$ and retain just the first order term,
\begin{equation}
\mathbf{U}(t)=\mathbf{U}^{(0)}(t)+\varepsilon\mathbf{U}^{(1)}(t)
\label{expa}\end{equation}
In particular, from Eq.~(\ref{defT}), we obtain
\begin{equation}
2\pi=\Phi^{(0)}(T)+\varepsilon\Phi^{(1)}(T)=\Phi^{(0)}(T^{(0)})+\varepsilon\Phi^{(1)}(T^{(0)})+
\varepsilon\frac{\partial\Phi^{(0)}}{\partial t}(T^{(0)})T^{(1)}
\end{equation}
where we have expanded $T$ as well, assuming that
$T=T^{(0)}+\varepsilon T^{(1)}$.
Since, $\Phi^{(0)}=2\pi$, we conclude that
\begin{equation}
T^{(1)}=-\frac{\Phi^{(1)}(T^{(0)})}{f_{1}}
\label{def2T}\end{equation}
where $f_{1}=\frac{\partial\Phi^{(0)}(T^{(0)})}{\partial t}$ is
determined by the right-hand side of (\ref{eqtot}) with $\varepsilon=0$,
namely, it is the first component of $\mathbf{F}$.

It can be easily seen that the zeroth and first order components satisfy
the differential equations
\begin{eqnarray}
\dot{\mathbf{U}}^{(0)} & = & \mathbf{F}(\mathbf{U}^{(0)})\label{eq:unpert}\\
\dot{\mathbf{U}}^{(1)} & = & \mathbf{DF}(\mathbf{U}^{(0)})\mathbf{U}^{(1)}+
\mathbf{G}(\mathbf{U}^{(0)},\Omega t+\psi)
\label{eq:genvar}\end{eqnarray}
where $\mathbf{DF}$ denotes the Jacobian of the velocity field $\mathbf{F}$
and we have introduced an explicit dependence on the phase $\psi$,
as it changes in going from one to the next section. The equation
for the first order correction can be formally solved,
\begin{equation}
\mathbf{U}^{(1)}=\int_{0}^{T^{(0)}}d\tau\mathbf{W}(T^{(0)},\tau)\mathbf{G}(\mathbf{U}^{(0)}
(\tau),\Omega\tau+\psi)
\label{eq:U1}\end{equation}
where $\mathbf{W}(t,\tau)$ is the matrix of fundamental solutions
of the equation $\dot{\mathbf{U}}=\mathbf{DF}(\mathbf{U}^{(0)})\mathbf{U}$.
Since $\mathbf{G}$ contains only first harmonics in $\psi$, it can
be decomposed into sine and cosine components,
\begin{equation}
\mathbf{G}(\mathbf{U}^{(0)},\Omega\tau+\psi)=\mathbf{G}^{c}(\mathbf{U}^{(0)},
\Omega\tau)\cos\psi+\mathbf{G}^{s}(\mathbf{U}^{(0)},\Omega\tau)\sin\psi
\end{equation}
where
\begin{equation}
\mathbf{G}^{c}(\mathbf{U}^{(0)},\Omega\tau)=\left(\begin{matrix}-\sin^{2}\Phi^{(0)}
   \cos\Omega\tau-\cos^{2}\Phi^{(0)}\sin\Omega\tau\\
R^{(0)}\sin2\Phi^{(0)}\cos(\Omega\tau+\pi/4)/\sqrt{2}\\
0\end{matrix}\right)
\end{equation}
and
\begin{equation}
\mathbf{G}^{s}(\mathbf{U}^{(0)},\Omega\tau)=\left(\begin{matrix}\sin^{2}\Phi^{(0)}
   \sin\Omega\tau-\cos^{2}\Phi^{(0)}\cos\Omega\tau\\
-R^{(0)}\sin2\Phi^{(0)}\sin(\Omega\tau+\pi/4)/\sqrt{2}\\
0\end{matrix}\right)
\end{equation}

Accordingly, as it follows from (\ref{eq:U1}), $\mathbf{U}^{(1)}$ can
be decomposed in the same way
\begin{equation}
\mathbf{U}^{(1)}=\mathbf{M}^{c}(r,z)\cos\psi+\mathbf{M}^{s}(r,z)\sin\psi
\label{decom}\end{equation}
where
\begin{equation}
\mathbf{M}^{c}(r,z)=\int_{0}^{T^{(0)}}d\tau\mathbf{W}(T^{(0)},\tau)
 \mathbf{G}^{c}(\mathbf{U}^{(0)}(\tau),\Omega\tau)
\end{equation}
and a similar equation holds for $\mathbf{M}^{s}(r,z)$.

The first component of Eq.~(\ref{decom}) gives $\Phi^{(1)}$. After
substituting it into Eq.~(\ref{def2T}), we obtain,
\begin{equation}
T=T^{(0)}(r,z)-\frac{\varepsilon}{f_{1}}(M_{1}^{c}(r,z)\cos\psi+M_{1}^{s}(r,z)
\sin\psi)
\end{equation}
where the subscripts indicate once more the component of the vector.
Accordingly, the new phase $\psi'$ is
\begin{equation}
\psi'=\psi+\Omega T=\psi+\Omega T^{(0)}(r,z)-\frac{\varepsilon\Omega}{f_{1}}[M_{1}^{c}(r,z)
\cos\psi+M_{1}^{s}(r,z)\sin\psi]
\end{equation}
On the other hand, from the second and third components of Eq.~(\ref{expa})
we obtain the new values $r'$ and $z'$ by also expanding the expression
of $T$ around $T^{(0)}$,
\begin{equation}
\mathbf{U}(T)=\mathbf{U}^{(0)}(T^{(0)})+\varepsilon\mathbf{U}^{(1)}(T^{(0)})+
 \varepsilon\mathbf{F}(\mathbf{U}^{(0)})T^{(1)}
\end{equation}
Straightforward but tedious calculations lead to
\begin{eqnarray}
\psi' & = & \psi+\langle T^{(0)}\rangle\Omega+A_{1}+\varepsilon(B_{1}^{c}\cos\psi+B_{1}^{s}\sin\psi)\\
r' & = & A_{2}+\varepsilon(B_{2}^{c}\cos\psi+B_{2}^{s}\sin\psi)\\
z' & = & A_{3}+\varepsilon(B_{3}^{c}\cos\psi+B_{3}^{s}\sin\psi)
\end{eqnarray}
where $\langle T^{(0)}\rangle$ is the average period of the unperturbed 
R\"ossler oscillator. The functions
$A_{i}$ can be determined by integrating the unperturbed equations
\begin{eqnarray}
A_{1}(r,z) & = & \left[T^{(0)}(r,z)-\langle T^{(0)}\rangle\right]
\,\Omega\nonumber \\
A_{2}(r,z) & = & R^{(0)}(r,z;T^{(0)})\\
A_{3}(r,z) & = & Z^{(0)}(r,z;T^{(0)})\nonumber
\end{eqnarray}
while the functions $B^c_i$ read as
\begin{eqnarray}
B_{1}^{c}(r,z) & = & -\Omega M_{1}^{c}(r,z)/f_{1}\nonumber \\
B_{2}^{c}(r,z) & = & M_{2}^{c}(r,z)-f_{2}M_{1}^{c}/f_{1}\\
B_{3}^{c}(r,z) & = & M_{3}^{c}(r,z)-f_{3}M_{1}^{c}/f_{1}\nonumber
\end{eqnarray}
and similar equations hold for the sine components.

\section{Second order contribution to the Lyapunov vector}
\label{app2}
In this appendix we derive a closed expression for the term $I_2$, which
accounts for the contribution to the LE arising from second order-corrections
to the LV direction (see Eq.~\toref{eq:i2}). To this pourpose, we have to
consider all terms of order $\eps$ in Eq.~(\ref{eq:vec1}), starting from
$v'_1(r',\psi') \equiv v'_1(r'_0,\psi'_0) + \eps \delta v'_1$, with
\begin{eqnarray}
\delta v'_1 =  + bc \Big[ L(r'_0)\cos(\psi'_0+\alpha)
  - R(r'_0)\sin(\psi'_0+\alpha) \Big] \cos \psi 
    \nonumber \\
 + 2c^2 g(r) \Big[ L_r(r'_0) \sin(\psi'_0+\alpha)
 + R_r(r'_0)\cos(\psi'_0+\alpha) \Big] \cos(\psi+\alpha)
\end{eqnarray}
where $r'_0 = f(r)$ and $\psi'_0 = \psi + K + \Delta r$ are the iterates of the
unperturbed GSF as defined in section \ref{measure_lyap}. In the following we
shall not care about the possible lack of differentiability along the
direction $r$ for two reasons: {\it i)} we have verified that setting up a more
accurate procedure leads to the same results, but its presentation would 
be more cumbersome; {\it ii)} the procedure is in itself correct, because in
the end we are interested in the integral that is insensitive to the presence of
singularities.

The recursive equation for the second order term writes as
\begin{equation}
v_2(r,\psi) =  \frac{1}{f_r(r)} v'_2(r'_0,\psi'_0) + s(r,\psi)
\label{eq:v2}
\end{equation}
where we have introduced the source term
\begin{equation} 
 s(r,\psi) =  \frac{1}{f_r(r)} \Big[-2c g_r(r) \cos(\psi + \alpha) v_1(r,\psi) +
\delta v'_1 +  v'_1(r'_0,\psi'_0) (\Delta v_1 - b \sin \psi) \Big]
\end{equation}
and $v_1'$ is given by Eq.~\toref{eq:rec}.

Being interested in the integral $I_2$ (see Eq.~(\ref{eq:i2})), we see that
the integration over $\psi$ can be easily performed since $\psi'_0$ ranges
over $[0,2\pi]$. More delicate is the integral over $r$ because of the folding
of $r'_0$. However, one can still solve the problem by separately integrating
over the negative and positive values of $r$, i.e. the two atoms of the finite
Markov partition. By introducing the integral over negative $r$-values 
\begin{equation}
I_2^- = \int_{-1}^0 dr \int_0^{2\pi} d\psi \, v_2(r,\psi)
\end{equation}
and analogously definining $I_2^+$, we obtain from Eq.~(\ref{eq:v2})
\begin{eqnarray}
I_2^- &=&  \frac{I_2^-+I_2^+}{4} + S \nonumber \\
I_2^+ &=&  -\frac{I_2^-+I_2^+}{4} + S 
\label{eq:int}
\end{eqnarray}
where $S$ is the integral of $s(r,\psi)$. 
\beq
S = \int_{-1}^1 dr \int_0^{2\pi} d\psi \, s(r,\psi)
\eeq
We thus eventually obtain
\begin{equation}
I_2 = I_2^- + I_2^+ = S .
\end{equation}
It is now convenient to express $s(r,\psi)$ as a function of $v_1'$ only, 
\begin{equation} 
 s(r,\psi) =  \frac{1}{f_r(r)} \Big\{ - bv_1' \sin \psi +  \delta v'_1  +
 \frac{\Delta v_1' - 2c g_r \cos(\psi +\alpha)} {f_r(r)} [v_1' + 2c g(r)
 \sin(\psi +\alpha)] \Big\}
\end{equation}
Upon integrating over $\psi$, we obtain,
\begin{eqnarray}
S =  \pi c^2\frac{\Delta}{4} \int_{-1}^1 dy [ L^2(y) + R^2(y)] \nonumber \\
-2 \pi c^2 \int_{0}^1 dr \, (r^2-r) \sin (\Delta r) \Big[ L_r(1-2r) \cos K
 - R_r(1-2r)  \sin K \Big]  \nonumber \\ 
+ \pi c^2 \Delta \int_{0}^1 dr \, (r^2-r) \cos(\Delta r) \Big[ L(1-2r) \cos K
 - R(1-2r) \sin K \Big] \nonumber \\ 
+ \pi c^2 \int_{0}^1 dr \, (1-2r) \sin(\Delta r) \Big[ L(1-2r) \cos K
 - R(1-2r) \sin K \Big]
\label{eq:stot}
\end{eqnarray}
After integrating by parts the integral involving $L_r$ and $R_r$ and
rescaling the dummy variable, we finally arrive at the desired result:
\begin{eqnarray}
I_2 &=&  \pi c^2\frac{\Delta}{4} \int_{-1}^1 dy [ L^2(y) + R^2(y)] + \\
&&\pi c^2 \int_{-1}^1 dy \sin \frac{\Delta(1-y)}{2}  y
  \Big[ L(y) \cos K  - R(y) \sin K \Big]
\label{eq:stot2}
\end{eqnarray}

\end{document}